\documentclass[amsmath,amssymb,a4paper,prc,preprint,superscriptaddress,showpacs]{revtex4}

\usepackage{bm,hyphenat,xspace}
\usepackage{graphicx,epsfig}

\newcommand{\Eq}{Eq.}

\newcommand{\Ref}{Ref.}
\newcommand{\Refs}{Refs.}
\newcommand{\Sect}{Sec.}

\newcommand {\mbf}[1]{{\mathbf{#1}}}

\newcommand{\fm}{\:\mathrm{fm}}
\newcommand{\cm}{\mathrm{c\!\:\!.m\!\:\!.}}
\newcommand{\He}{{}^3\mathrm{He}}
\newcommand{\zr}{\mathcal{Z}_R^{-\frac12}}
\newcommand{\zR}{\mathcal{Z}_R}

\begin{document}

\title{Benchmark calculation for proton-deuteron elastic scattering 
observables including Coulomb}

\author{A. Deltuva} 
\email{deltuva@cii.fc.ul.pt}
\thanks{on leave from Institute of Theoretical Physics and Astronomy,
Vilnius University, Vilnius 2600, Lithuania}
\affiliation{Centro de F\'{\i}sica Nuclear da Universidade de Lisboa, 
P-1649-003 Lisboa, Portugal }

\author{A. C. Fonseca} 
\affiliation{Centro de F\'{\i}sica Nuclear da Universidade de Lisboa, 
P-1649-003 Lisboa, Portugal }

\author{A. Kievsky}
\affiliation{Istituto Nazionale di Fisica Nucleare, 
I-56100 Pisa, Italy }
\affiliation{Dipartimento di Fisica, Universita' di Pisa,
I-56100 Pisa, Italy }

\author{S. Rosati}
\affiliation{Istituto Nazionale di Fisica Nucleare, 
I-56100 Pisa, Italy }
\affiliation{Dipartimento di Fisica, Universita' di Pisa,
I-56100 Pisa, Italy }

\author{P. U. Sauer}
\affiliation{Institut f\"ur Theoretische Physik,  Universit\"at Hannover,
  D-30167 Hannover, Germany}

\author{M. Viviani}
\affiliation{Istituto Nazionale di Fisica Nucleare, 
I-56100 Pisa, Italy }
\affiliation{Dipartimento di Fisica, Universita' di Pisa,
I-56100 Pisa, Italy }
\received{March 4, 2005}


\begin{abstract}
Two independent calculations of proton-deuteron elastic scattering observables
including Coulomb repulsion between the two protons are compared in the proton
lab energy region between 3 MeV and 65 MeV. The hadron dynamics
is based on the purely nucleonic charge-dependent AV18 potential.
Calculations are done both in coordinate space and momentum space. 
The coordinate-space calculations are based on a
variational solution of the three-body Schr\"odinger equation using 
a correlated hyperspherical expansion for the wave function.
The momentum-space calculations proceed via the solution of the 
Alt-Grassberger-Sandhas equation using the screened Coulomb potential 
and the renormalization approach. 
Both methods agree within 1\% on all observables, showing the reliability 
of both numerical techniques in that energy domain. 
At energies below three-body breakup threshold
the coordinate-space method remains favored whereas at energies higher than
65 MeV the momentum-space approach seems to be more efficient.   
\end{abstract} 
  

\maketitle

\section{Introduction \label{In}} 
\bigskip 

Although there is a long history of theoretical work on the solution of
the Coulomb problem in three-particle scattering
\cite{alt:78a,berthold:90a,kievsky:96a,kievsky:01a,chen:01a,alt:02a,
suslov:04a}, the work of \Refs~\cite{kievsky:96a,kievsky:01a} 
pioneered the effort on fully converged numerical calculations for 
proton-deuteron $(pd)$ elastic scattering including
the Coulomb repulsion between protons together with realistic nuclear
interactions. In their work the authors use the charge-dependent AV18
potential  together with the Urbana IX three-nucleon force and
proceed to solve the three-particle Schr\"odinger  equation using the Kohn
variational principle (KVP); the wave function satisfies appropriate Coulomb
distorted asymptotic boundary conditions and is expanded at short distances
in a pair correlated hyperspherical harmonics basis set. The results
presented were fully converged vis-\`a-vis the size of the basis set and the
angular momentum states included in the calculation. 
In parallel a benchmark calculation was performed~\cite{kievsky:01b} 
where results obtained variationally
were compared with those obtained from the solution of coordinate-space
Faddeev equations for the AV14 potential 
at energies below three-body breakup threshold. 

In a recent publication~\cite{deltuva:05a} the momentum-space solution of the
Alt-Grassberger-Sandhas (AGS) equation~\cite{alt:67a} for two protons
and a neutron was successfully applied, not only to $pd$ elastic
scattering but also to radiative $pd$ capture and two-body
electromagnetic disintegration of $\He$. 
The treatment of the Coulomb interaction is based on the ideas
proposed by Taylor~\cite{taylor:74a} for two charged particle scattering and
extended in \Ref~\cite{alt:78a} for three-particle scattering with
two charged particles alone. The Coulomb potential is screened, standard
scattering theory for short-range potentials is used, and the obtained
results are corrected for the unscreened limit using the renormalization
prescription~\cite{taylor:74a,alt:78a}. The results presented in
\Ref~\cite{deltuva:05a} are converged vis-\`a-vis the screening radius $R$
and the number of included two-body and three-body angular momentum states.
Although in \Ref~\cite{deltuva:05a} the hadron dynamics is based 
on the purely nucleonic charge-dependent (CD) Bonn potential 
and its realistic coupled-channel extension 
CD Bonn + $\Delta$, allowing for single virtual 
$\Delta$-isobar excitation, 
other realistic potential models may be used easily as well.  
 
Motivated by recent experimental efforts in the measurements of $pd$ elastic
observables~\cite{sagara:94a,exp1,exp2}, 
in the present paper we present benchmark
results for a number of $pd$ elastic scattering observables, 
both below and above three-body breakup threshold, using
the charge-dependent AV18~\cite{wiringa:95a} two-nucleon potential
and no three-nucleon force.
In \Sect~\ref{Methods} we make a short 
description of the methods we use, in \Sect~\ref{sec:results} we present 
the results, and  in \Sect~\ref{sec:conclusions} the conclusions.

\section{The Methods \label{Methods}}

In this section we briefly introduce both methods and provide the basic 
framework for a general understanding of the technical procedures;
further details may be found in the appropriate references.
We choose to describe the method based on KVP using its traditional notation,
which we attempt to carry over to the discussion of the integral equation
approach in \Sect~\ref{sec:IE} and \ref{sec:results}.
Therefore the presentation of the integral equation approach will not be 
in the notation used in \Ref~\cite{deltuva:05a}.

\subsection{The Kohn variational principle}

\def\x{{\bf x}}
\def\y{{\bf y}}
\def\r{{\bf r}}
\def\a{{\alpha}}
\def\r{{\rho}}
\def\aa{{\alpha\alpha'}}
\def\kk{{kk'}}
\def\ra{\rightarrow}
\def\htm{{\hbar^2\over M_N}}

The KVP can be used to describe nucleon-deuteron $(Nd)$ elastic scattering.
Below the three-body breakup threshold the
collision matrix is unitary and the problem can be formulated
in terms of the real reactance matrix ($K$--matrix).
Above the three-body breakup threshold the elastic part of the 
collision matrix is no longer unitary and the formulation in terms of the
$S$-matrix, the complex form of the KVP, is convenient.
Referring to Refs.~\cite{kievsky:01a,KRV99,VKR00} for details, 
a brief description of the method is given below. 
The scattering wave function (w.f.) $\Psi$ is
written as sum  of two terms
 \begin{equation}
   \Psi=\Psi_C+\Psi_A \ 
\label{eq:psi}
 \end{equation}
which carry the appropriate asymptotic boundary conditions.
The first term, $\Psi_C$, describes the
system when the three--nucleons are close to each other. For large
interparticle separations and energies below the
three-body breakup threshold it goes to zero, 
whereas for higher energies it must
reproduce a three outgoing particle state. It
is written as a sum of three Faddeev--like amplitudes
corresponding to the three cyclic permutations of the particle indices.
Each amplitude $\Psi_C(\x_i,\y_i)$, where $\x_i,\y_i$ are
the Jacobi coordinates corresponding to the $i$-th permutation, has
total angular momentum $JJ_z$ and total isospin $TT_z$ and is
decomposed into channels using $LS$ coupling, namely
\begin{eqnarray}
     \Psi_C(\x_i,\y_i) &=& \sum_{\alpha=1}^{N_c} \phi_\alpha(x_i,y_i)
     {\cal Y}_\alpha (jk,i)  \\
     {\cal Y}_\alpha (jk,i) &=&
     \Bigl\{\bigl[ Y_{\ell_\alpha}(\hat x_i)  Y_{L_\alpha}(\hat y_i)
     \bigr]_{\Lambda_\alpha} \bigl [ s_\alpha^{jk} s_\alpha^i \bigr ]
     _{S_\alpha}
      \Bigr \}_{J J_z} \; \bigl [ t_\alpha^{jk} t_\alpha^i \bigr ]_{T T_z},
\end{eqnarray}
where $x_i,y_i$ are the moduli of the Jacobi coordinates and
${\cal Y}_\alpha$ is the angular-spin-isospin function for each channel.
The maximum number of channels considered in the expansion is $N_c$.
The two-dimensional amplitude $\phi_\alpha$ is expanded in terms of the
pair correlated hyperspherical harmonic basis \cite{KVR93,KVR94}
\begin{equation}
     \phi_\alpha(x_i,y_i) = \rho^{-5/2} f_\alpha(x_i)
     \left[ \sum_K u^\alpha_K(\rho) {}^{(2)}P^{\ell_\alpha,L_\alpha}_K(\phi_i)
     \right] \ ,
\label{eq:PHH}
\end{equation}
where the hyperspherical variables, the hyperradius $\rho$ and
the hyperangle $\phi_i$, are defined by the relations
$x_i=\rho\cos{\phi}_i$ and $y_i=\rho\sin{\phi}_i$. The factor
${}^{(2)}P^{\ell,L}_K(\phi)$ is a hyperspherical polynomial and
$f_\alpha(x_i)$ is a pair correlation function introduced
to accelerate the convergence of the expansion. For small values
of the interparticle distance $f_\alpha(x_i)$ is regulated by the
$NN$ interaction whereas for large separations 
$f_\alpha(x_i)\rightarrow 1$.

The second term, $\Psi_A$, in the variational wave function of
Eq.(\ref{eq:psi})
describes the asymptotic  motion of a deuteron relative to the third
nucleon. It can also be written  as a sum
of three amplitudes with the generic one having the form
\begin{equation}
   \Omega^\lambda_{LSJ}(\x_i,\y_i) = \sum_{l_\a=0,2} w_{l_\a}(x_i)
       {\cal R}^\lambda_L (y_i)
       \left\{\left[ [Y_{l_\a}({\hat x}_i) s_\a^{jk}]_1 s^i \right]_S
        Y_L({\hat y}_i) \right\}_{JJ_z}
       [t_\a^{jk}t^i]_{TT_z}\ , \label{eq:omega}
\end{equation}
where $w_{l_\a}(x_i)$ is the deuteron w.f. radial component in the 
state $l_\a =0,2$.
In addition, $s_\a^{jk}=1,t_\a^{jk}=0$ and $L$ is the relative nucleon-deuteron
angular momentum. The superscript $\lambda$ indicates
the regular ($\lambda\equiv R$) or the irregular ($\lambda\equiv I$)
solution. In the $pd$ $(nd)$ case, the functions
${\cal R}^\lambda$ are related to
the regular or irregular  Coulomb (spherical Bessel) functions.
The functions $\Omega^\lambda$ can be combined to form a general
asymptotic state ${}^{(2S+1)}L_J$
\begin{equation}
\Omega^+_{LSJ}(\x_i,\y_i) =  \Omega^0_{LSJ}(\x_i,\y_i)+
 \sum_{L'S'}{}^J{\cal L}^{SS'}_{LL'}\Omega^1_{L'S'J}(\x_i,\y_i)  \ ,
\end{equation}
where
\begin{eqnarray}
\Omega^0_{LSJ}(\x_i,\y_i) =& u_{00}\Omega^R_{LSJ}(\x_i,\y_i)+
                            u_{01}\Omega^I_{LSJ}(\x_i,\y_i) \ , \\
\Omega^1_{LSJ}(\x_i,\y_i) =& u_{10}\Omega^R_{LSJ}(\x_i,\y_i)+
                            u_{11}\Omega^I_{LSJ}(\x_i,\y_i)  \ .
\end{eqnarray}
The matrix elements $u_{ij}$ can be selected according to the
four different choices of the matrix ${\cal L}=$ $K$-matrix,
$K^{-1}$-matrix, $S$-matrix or $T$-matrix. A general
three-nucleon scattering w.f. for an incident
state with relative angular momentum $L$, 
spin $S$ and total angular momentum $J$ is
\begin{equation}
\Psi^+_{LSJ}=\sum_{i=1,3}\left[ \Psi_C(\x_i,\y_i)+\Omega^+_{LSJ}(\x_i,\y_i)
             \right] \ ,
\end{equation}
and its complex conjugate is $\Psi^-_{LSJ}$. A variational estimate of the
trial parameters in the w.f. $\Psi^+_{LSJ}$ can
be obtained by requiring, in accordance with
the generalized KVP, that the functional
\begin{equation}
[{}^J{\cal L}^{SS'}_{LL'}]= {}^J{\cal L}^{SS'}_{LL'}-{\frac{2}{{\rm det}(u)}}
\langle\Psi^-_{LSJ}|H-E|\Psi^+_{L'S'J}\rangle \ ,
\label{eq:kohn}
\end{equation}
be stationary. Below the three-body breakup threshold,
due to the unitarity of the
$S$-matrix, the four forms for the ${\cal L}$-matrix are equivalent.
However, it was shown that when the
complex form of the principle is used, there is a considerable
reduction of numerical instabilities~\cite{kiev97}. 
Above the three-body breakup threshold it is convenient to formulate 
the variational principle
in terms of the $S$--matrix. Accordingly, we get the following functional:
\begin{equation}
[{}^J{S}^{SS'}_{LL'}]= {}^J{S}^{SS'}_{LL'}+{i}
\langle\Psi^-_{LSJ}|H-E|\Psi^+_{L'S'J}\rangle \ .
\label{eq:ckohn}
\end{equation}

The variation of the functional with respect to the hyperradial
functions $u^\alpha_K(\rho)$
leads to the following set of coupled equations:
\begin{equation}
  \sum_{\alpha',k'}
  \Bigl[ A^\aa_\kk (\r ){d^2\over d\r^2}+ B^\aa_\kk (\r ){d\over d\r}
        + C^\aa_\kk (\r )+
     {M_N\over\hbar^2} E\; N^\aa_\kk (\r )\Bigr ]
        u^{\alpha'}_{k'}(\r)= D^\lambda_{\alpha k}(\rho) \ .
   \label{eq:siste}
\end{equation}
For each asymptotic state $^{(2S+1)}L_J$  two different inhomogeneous terms
are constructed corresponding to the asymptotic $\Omega^\lambda_{LSJ}$
functions with $\lambda\equiv 0,1$. Accordingly, two sets of solutions
are obtained and
combined to minimize the functional (\ref{eq:ckohn}) with respect to
the $S$-matrix elements. This is the first order solution, the second order
estimate of the $S$-matrix is obtained after replacing the first order
solution in Eq.(\ref{eq:ckohn}).

In order to solve the above system of equations appropriate
boundary conditions must be specified for the hyperradial functions.
For energies below the three-body breakup threshold
they go to zero when $\rho\rightarrow\infty$, whereas
for higher energy they asymptotically describe the breakup configuration.
The boundary conditions to be applied in this case have
been discussed in Refs.~\cite{kievsky:01a,VKR00,KVR97} 
and are briefly illustrated below.
To simplify the notation let us label the basis
elements with the index $\mu\equiv[\a,K]$,
and introduce the completely antisymmetric correlated
spin-isospin-hyperspherical basis element ${\cal Q}_\mu(\rho,\Omega)$
as linear combinations of the products
\begin{equation}
\label{eq:bco}
     \sum_{i=1}^3
      f_\a(x_i)\; {}^{(2)}P^{\ell_\alpha,L_\alpha}_K(\phi_i)
      {\cal Y}_\alpha(jk,i) \ ,
\end{equation}
which depend on $\rho$ through the correlation factor. 
In terms of the ${\cal Q}_\mu(\rho,\Omega)$ the internal part is written as
\begin{equation}
   \Psi_C= \rho^{-5/2}\sum_{\mu=1}^{N_m}
    \omega_{\mu}(\rho) {\cal Q}_\mu(\rho,\Omega) \ ,
\end{equation}
with $N_m$ the total number of basis functions considered.
The hyperradial functions $u_\mu(\rho)$ and $\omega_{\mu}(\rho)$
are related by an unitary transformation imposing that
the ``uncorrelated'' basis elements ${\cal Q}^0_\mu(\Omega)$,
obtained by setting all the correlation functions $f_\a(x_i)=1$, 
form an orthogonal basis. 
Explicitly, the matrix elements of the norm $N$ behave as
\begin{equation}
   N_{\mu\mu'}(\rho)= \int d\Omega\;
     {\cal Q}_\mu(\rho,\Omega)^\dag
     {\cal Q}_{\mu'} (\rho,\Omega) \ra N^{(0)}_{\mu\mu'}
    +{  N^{(3)}_{\mu\mu'}\over \rho^3}+{\cal O}(1/\rho^5)
      \ ,\qquad {\rm for\ }\rho\ra\infty\ ,  \label{eq:n}
\end{equation}
where, in particular,
\begin{equation}
   N^{(0)}_{\mu\mu'}= \int d\Omega\;
      {\cal Q}^0_\mu (\Omega)^\dag
      {\cal Q}^0_{\mu'} (\Omega)\ .  \label{eq:n1}
\end{equation}
is diagonal with diagonal elements ${\cal N}_\mu$
either $1$ or $0$. Therefore, some correlated elements
have the property: ${\cal Q}_\mu(\rho,\Omega) \ra 0$ as $\rho\ra\infty$.
In the following we arrange the new basis in such a way that for values
of the index $\mu\le\overline{N}_m$ the eigenvalues of the
norm are ${\cal N}_\mu=1$ and
for $\overline{N}_m+1 \le \mu \le N_m$ they are ${\cal N}_\mu=0$.
 
For $\rho\ra\infty$, neglecting terms going to zero faster
than $\rho^{-2}$, the asymptotic expression of the set of Eqs.(\ref{eq:siste})
rotated using the unitary transformation defined above,
reduces to the form
\begin{equation}
 \label{eq:c0}
 \sum_{\mu'} \biggl\{
-\htm  \left( {d^2\over d\rho^2} -{{\cal K}_\mu({\cal K}_\mu +1)\over\rho^2}
 + Q^2 \right ){\cal N}_\mu
 \delta_{\mu,\mu'} +
   {2\;Q\; \chi_{\mu\mu'}\over \rho} \;
   +{\cal O}({1\over\rho^3})\biggr\}\omega_{\mu'}(\rho) = 0 \ ,
\end{equation}
where $E=\hbar^2 Q^2/M_N$ and ${\cal K}_\mu= G_\mu+3/2$. Here $G_\mu$ is
the grand-angular quantum number defined as $G_\mu=l_\alpha+L_\alpha + 2 K$ 
and the matrix $\chi$ is defined as
\begin{equation}
\label{eq:c}
   { \chi}_{\mu\mu'}= \int d\Omega\;
      {\cal Q}^0_{\mu} (\Omega)^\dag
     \; \hat \chi \;
      {\cal Q}^0_{\mu'} (\Omega)
      \ .
\end{equation}
The dimensionless operator $\hat\chi$ originates from the Coulomb interaction
as
\begin{equation}
   \hat \chi = {M_N\over 2\hbar^2 Q}
  \sum_{i=1}^3 {e^2\over \cos\phi_i} {1+\tau_{j,z} \over 2}
  {1+\tau_{k,z} \over 2} \ .
\label{eq:chi}
\end{equation}
It should be noticed that $\chi_{\mu\mu'}=0$ if $\mu,\mu'>{\overline{N}_m}$.

In practice,
the functions $\omega_\mu(\rho)$ are chosen to be regular at the origin, i.e.
$\omega_\mu(0)=0$ and, in accordance with the equations to be satisfied for
$\rho\ra\infty$, to have the following behavior
($\mu\le\overline{N}_m$)
\begin{equation}
\label{eq:asy2}
  \omega_\mu(\rho) \rightarrow
   - \sum_{\mu'=1}^{\overline{N}_m}
  \left ( e^{-i {\hat \chi} \ln 2 Q\rho} \right)_{\mu\mu'}\;
   b_{\mu'} \; e^{i Q\rho} \ ,
\end{equation}
where $ b_{\mu'}$ are unknown coefficients. This form corresponds
to the asymptotic behavior of three outgoing particles
interacting through the Coulomb potential~\cite{merkuriev2}.
In the case of $nd$ scattering ($\chi\equiv 0$)
the outgoing solutions evolve as outgoing Hankel functions
$H^{(1)}(Q\rho)$ ($\omega_\mu(\rho)\rightarrow -b_\mu e^{iQ\rho}$).

For values of the index $\mu > \overline{N}_m$ the eigenvalues of the
norm are ${\cal N}_\mu=0$ and the leading terms
in Eq.(\ref{eq:c0}) vanish. So, the asymptotic behavior of these
$\omega_\mu$ functions is governed by the next order terms.
However, for $\mu > \overline{N}_m$, it is verified that
$\omega_\mu{\cal Q}_\mu\rightarrow 0$ as $\rho\rightarrow\infty$.

In order to solve the system of Eqs.(\ref{eq:siste})
the hyperradial functions are expanded in terms of Laguerre
polynomials plus an auxiliary function
\begin{equation} \label{eq:M}
 \omega_\mu(\rho)=\rho^{5/2}\sum_{m=0}^M A^m_{\mu} 
L^{(5)}_m(z)\exp(-{z\over 2}) +A^{M+1}_{\mu} \overline \omega_{\mu}(\rho) \ ,
\end{equation}
where $z=\gamma\rho$ and $\gamma$ is a nonlinear parameter.
The linear parameters $A^m_{\mu}$ $(m=0,....,M+1)$
are determined by the variational procedure.
 
The inclusion of the auxiliary functions $\overline \omega_{\mu}(\rho)$
defined in Eq.(\ref{eq:M}) is useful for reproducing the oscillatory
behavior shown by the hyperradial functions for $\rho\gtrsim 30$ fm.
Otherwise
a rather large number $M$ of polynomials should be included in
the expansion. A convenient choice is to take them
as the regular solutions of a one dimensional
differential equation corresponding to the $\mu$-th equation of
the system whose asymptotic behavior is the one of Eq.(\ref{eq:c0}).
In the cases considered here
the solutions obtained for the $S$-matrix stabilize
for values of the matching radius $\rho_0>100$ fm.

\subsection{The integral equation approach  \label{sec:IE}}

The integral equation to be solved is the AGS equation~\cite{alt:67a}
for three-particle scattering where each pair of nucleons
interacts through the strong potential $v$ and the Coulomb potential
$w$ acts only between charged nucleons. The work in
\Ref~\cite{deltuva:05a} follows the seminal work of 
\Refs~\cite{taylor:74a,alt:78a} in the sense that the treatment of the 
Coulomb interaction is based
on screening, followed by the use of standard scattering theory for
short-range potentials and renormalization of the obtained results in order
to correct for the unscreened  limit. Nevertheless there are important
differences relative to \Ref~\cite{alt:02a} that are paramount to the fast
convergence of the calculation in terms of screening radius R and the
effective use of realistic interactions:

{\bf a)} We work with a screened Coulomb potential 
\begin{gather}
 w_R(r) = w(r) \; e^{-(r/R)^n}
\end{gather}
where $w (r) = \frac{\alpha}{r}$ is the true  Coulomb
potential, $\alpha$ being the fine structure constant and $n$ a power
controlling the smoothness of the screening. We prefer to work with a
sharper screening than the Yukawa screening $(n=1)$ of \Ref~\cite{alt:02a}
because we want to ensure that the screened Coulomb potential $w_R$
approximates well the true Coulomb one $w$ for distances $r<R$ and
simultaneously vanishes rapidly for $r>R$, providing a comparatively rapid
convergence of the partial wave expansion. In contrast, the sharp cutoff 
$(n \to \infty)$ yields unpleasant oscillatory behavior in momentum space
representation, leading to convergence problems. We find values
$3 \le n \le 6$ to provide a sufficient smoothness and fast convergence;
$n = 4$ is used for the calculations of this paper.

{\bf b)} Although the choice of the screened potential  improves the
partial wave convergence, the practical implementation of the solution of
AGS equation still places a technical  difficulty, i.e., the calculation of
the AGS operators for nuclear plus screened Coulomb potentials requires
two-nucleon partial waves with pair orbital angular momentum considerably
higher than required for the hadronic potential alone. In this context the
perturbation theory for higher two-nucleon partial waves developed in
Ref.~\cite{deltuva:03b} is a very efficient and reliable technical tool for
treating the screened Coulomb interaction in high partial waves.

As a result of these two technical implementations,  the
method~\cite{deltuva:03a} that was developed before for solving
three-particle AGS equations without Coulomb could be successfully used in
the presence of screened Coulomb. Using the usual three-body notation, the
full multichannel transition matrix  reads
\begin{subequations}\label{eq:a+b} 
\begin{gather} \label{eq:Uba}
  \begin{align} 
     U^{(R)}_{\beta \alpha}(Z) = {} & \bar{\delta}_{\beta \alpha} G_0^{-1}(Z)
     + \sum_{\sigma} \bar{\delta}_{\beta \sigma} T^{(R)}_\sigma (Z) G_0(Z) 
     U^{(R)}_{\sigma \alpha}(Z), 
  \end{align} 
\end{gather}
\noindent where the superscript $(R)$ denotes the dependence on the screening 
radius $R$ of the Coulomb potential, $G_0(Z) = (Z - H_0)^{-1}$ the free
resolvent, $\bar{\delta}_{\beta \alpha} = 1- \delta_{\beta\alpha}$, and 
\begin{gather} \label{eq:TR}
\begin{align} 
   T^{(R)}_\alpha (Z) = {}& (v_\alpha + w_{\alpha R}) + 
     (v_\alpha + w_{\alpha R})  G_0(Z) T^{(R)}_\alpha (Z).
  \end{align}
\end{gather}
\end{subequations} 
The two-particle transition matrix $T^{(R)}_\alpha (Z)$ results  from
the nuclear interaction $v_{\alpha}$ between hadrons plus the screened
Coulomb $w_{\alpha R}$ between charged nucleons ($w_{\alpha R} = 0$ otherwise).
As expected the full multichannel transition matrix
$U^{(R)}_{\beta \alpha}(Z)$ must contain the pure Coulomb transition
matrix $T^{\cm}_{\alpha R} (Z)$ derived from the screened Coulomb
$W^{\cm}_{\alpha R}$ between the spectator proton and the center of mass
(c.m.) of the remaining neutron-proton $(np)$ pair in channel $\alpha$
\begin{gather} \label{eq:Tcm}
  T^{\cm}_{\alpha R} (Z) = W^{\cm}_{\alpha R} + 
  W^{\cm}_{\alpha R} G^{(R)}_{\alpha} (Z) T^{\cm}_{\alpha R} (Z),
\end{gather}
where  $W^{\cm}_{\alpha R} = 0$ for $n(pp) \; \;\alpha$ channels 
and $G^{(R)}_{\alpha}$ the channel resolvent 
\begin{gather} \label{eq:GRa}
  G^{(R)}_\alpha (Z) = (Z - H_0 - v_\alpha - w_{\alpha R})^{-1}.
\end{gather}
In a system of two charged particles and a neutral one,  when 
$w_{\alpha R} = 0$, $\; W^{\cm}_{\alpha R} \ne 0 $  and vice versa. 

As demonstrated in \Refs~\cite{alt:78a,deltuva:05a} the split of 
the multichannel transition matrix 
\begin{gather} \label{eq:GR3}
  U^{(R)}_{\beta \alpha}(Z) = \delta_{\beta\alpha} 
  T^{\cm}_{\alpha R}(Z) + [ U^{(R)}_{\beta \alpha}(Z) -
    \delta_{\beta\alpha} T^{\cm}_{\alpha R}(Z)]
\end{gather} 
into a long-range part $ \delta_{\beta \alpha} T^{\cm}_{\alpha R}(Z) $ 
and a Coulomb distorted short-range  part 
$[U^{(R)}_{\beta\alpha}(Z) - \delta_{\beta\alpha} T^{\cm}_{\alpha R}(Z)]$
is extremely convenient to recover the unscreened Coulomb limit.
According to \Refs~\cite{alt:78a,deltuva:05a} the full $pd$ transition 
amplitude $ \langle  \phi_\beta (\mbf{q}_f) \nu_{\beta_f} | U_{\beta \alpha}
|\phi_\alpha (\mbf{q}_i) \nu_{\alpha_i} \rangle $ for the initial
and final channel states with relative $pd$ momentum $\mbf{q}_i$ and 
$\mbf{q}_f$, $q_f = q_i$, energy $E_{\alpha}(q_i)$, and discrete
quantum numbers $\nu_{\alpha_i}$ and $\nu_{\beta_f}$,
is obtained via the renormalization of the on-shell
$U^{(R)}_{\beta \alpha}(Z)$ with $Z = E_{\alpha}(q_i) + i0$
 in the infinite $R$ limit. For the screened
Coulomb transition matrix $T^{\cm}_{\alpha R}(Z)$,
contained in $ U^{(R)}_{\beta \alpha}(Z)$, that limit can be carried out 
analytically, yielding the proper Coulomb transition amplitude
$\langle \phi_\beta (\mbf{q}_f) \nu_{\beta_f} |T^{\cm}_{\alpha C}
|\phi_\alpha (\mbf{q}_i) \nu_{\alpha_i} \rangle $ \cite{alt:78a,taylor:74a},
while the Coulomb distorted short-range part requires the explicit use of a 
renormalization factor,
\begin{gather} \label{eq:UC2}
  \begin{split}
    \langle  \phi_\beta (\mbf{q}_f) & \nu_{\beta_f} | U_{\beta \alpha}
    |\phi_\alpha (\mbf{q}_i) \nu_{\alpha_i} \rangle   \\ = {}&
    \delta_{\beta \alpha}
    \langle \phi_\beta (\mbf{q}_f) \nu_{\beta_f} |T^{\cm}_{\alpha C}
    |\phi_\alpha (\mbf{q}_i) \nu_{\alpha_i} \rangle  \\ & +
    \lim_{R \to \infty} \{ \zr(q_f)
    \langle \phi_\beta (\mbf{q}_f) \nu_{\beta_f} |
            [ U^{(R)}_{\beta \alpha}(E_\alpha(q_i) + i0)  \\ & -
              \delta_{\beta\alpha} T^{\cm}_{\alpha R}(E_\alpha(q_i) + i0)]
            |\phi_\alpha (\mbf{q}_i) \nu_{\alpha_i} \rangle
            \zr(q_i) \}.
  \end{split}
\end{gather}
The renormalization factor
\begin{subequations}
  \begin{gather}\label{eq:zrq}
    \zR(q) = e^{-2i \phi_R(q)},
  \end{gather}
  contains a phase $\phi_R(q)$ which,  though  independent of the
  $pd$ relative orbital momentum $L$ in the infinite $R$ limit, 
  is given by \cite{taylor:74a}
  \begin{gather}    \label{eq:phiRl}
    \phi_R(q) = \sigma_L(q) -\eta_{LR}(q), 
  \end{gather}
\end{subequations}
where $\eta_{LR}(q)$ is the diverging screened Coulomb phase
shift corresponding to standard boundary conditions, and $\sigma_L(q)$
the proper Coulomb phase referring to logarithmically distorted
Coulomb boundary conditions. The limit of the Coulomb distorted
short-range part of the multichannel transition matrix 
$[ U^{(R)}_{\beta\alpha}(Z) - \delta_{\beta\alpha} T^{\cm}_{\alpha R}(Z)]$ 
has to be performed
numerically but, due to its short-range nature, the limit is reached
with sufficient accuracy at finite screening radii $R$. Furthermore,
due to the choice of screening and perturbation technique to deal with
high angular momentum states, $[ U^{(R)}_{\beta \alpha}(Z) -
\delta_{\beta\alpha} T^{\cm}_{\alpha R}(Z)]$ is calculated through the
numerical solution of Eqs.~(\ref{eq:a+b}) and
(\ref{eq:Tcm}), using partial-wave expansion.

In actual calculations we use the isospin formulation and, therefore, the
nucleons are considered identical. Instead  of \Eq~\eqref{eq:Uba} we 
solve a symmetrized AGS equation
\begin{gather} \label{eq:UR}
  U^{(R)}(Z) = P G_0^{-1}(Z) + P T^{(R)}_{\alpha}(Z) G_0(Z) U^{(R)}(Z),
\end{gather}
$P$ being the sum of the two cyclic three-particle permutation operators,
and use a properly symmetrized $pd$ transition amplitude
\begin{gather}
  \begin{split}\label{eq:Uasym}
    \langle \phi_\alpha (\mbf{q}_f)  & \nu_{\alpha_f} |
    U |\phi_\alpha (\mbf{q}_i) \nu_{\alpha_i} \rangle  \\ = {} &
    \langle \phi_\alpha (\mbf{q}_f) \nu_{\alpha_f} |
    T^{\cm}_{\alpha C}|\phi_\alpha (\mbf{q}_i) \nu_{\alpha_i} \rangle
    \\ & +  \lim_{R \to \infty}
    \{ \zr(q_f) \langle \phi_\alpha (\mbf{q}_f) \nu_{\alpha_f}|
       [ U^{(R)}(E_\alpha(q_i) + i0)  \\ & -
         T^{\cm}_{\alpha R}(E_\alpha(q_i) + i0)]
       |\phi_\alpha (\mbf{q}_i) \nu_{\alpha_i} \rangle  \zr(q_i) \}
  \end{split}
  \end{gather}
for the calculation of observables.
For further technical details we refer to \Ref~\cite{deltuva:05a}.  
 
\section{Results  \label{sec:results}}
\bigskip 

In this section we compare numerical calculations for a number of
elastic observables performed using the KVP and the integral equation 
approach. Three different lab energies have been considered: $3$, $10$, and
$65$ MeV. The Coulomb effects are expected to be
sizable in most of the observables at the first two energies.
The two methods use a different scheme to construct the
scattering states with total angular momentum and parity $J^\pi$.
In the KVP the $LS$ coupling is used and channels are ordered by
increasing values of $\ell_\alpha + L_\alpha$. The expansion of
the scattering state is truncated at values 
$\ell_\alpha + L_\alpha=L_{max}+2$, where $L_{max}$ is the maximum value
of $L$ corresponding to the asymptotic states ${}^{(2S+1)}L_J$. 
In the integral equation approach the $jj$ scheme has been used. The
channels have been ordered for increasing values of the two-body angular
momentum $j$ and for the strong interaction
the maximum value $j_{max}=5$ has been considered 
for the first two energies whereas at $65$ MeV the value $j_{max}=6$ 
has been used; the screened Coulomb interaction is taken into
account up to $j_{max}=25$ as described in \Ref~\cite{deltuva:05a}.
Both numerical calculations presented here are 
converged relative to the number of three-body partial waves. In addition, 
the variational calculations are converged relative to the size of
the hyperspherical basis set and, in the integral equation approach, the
results are converged with respect to the screening radius $R$.

\begin{figure*}
\begin{center}
\includegraphics[width=17cm,height=19cm]{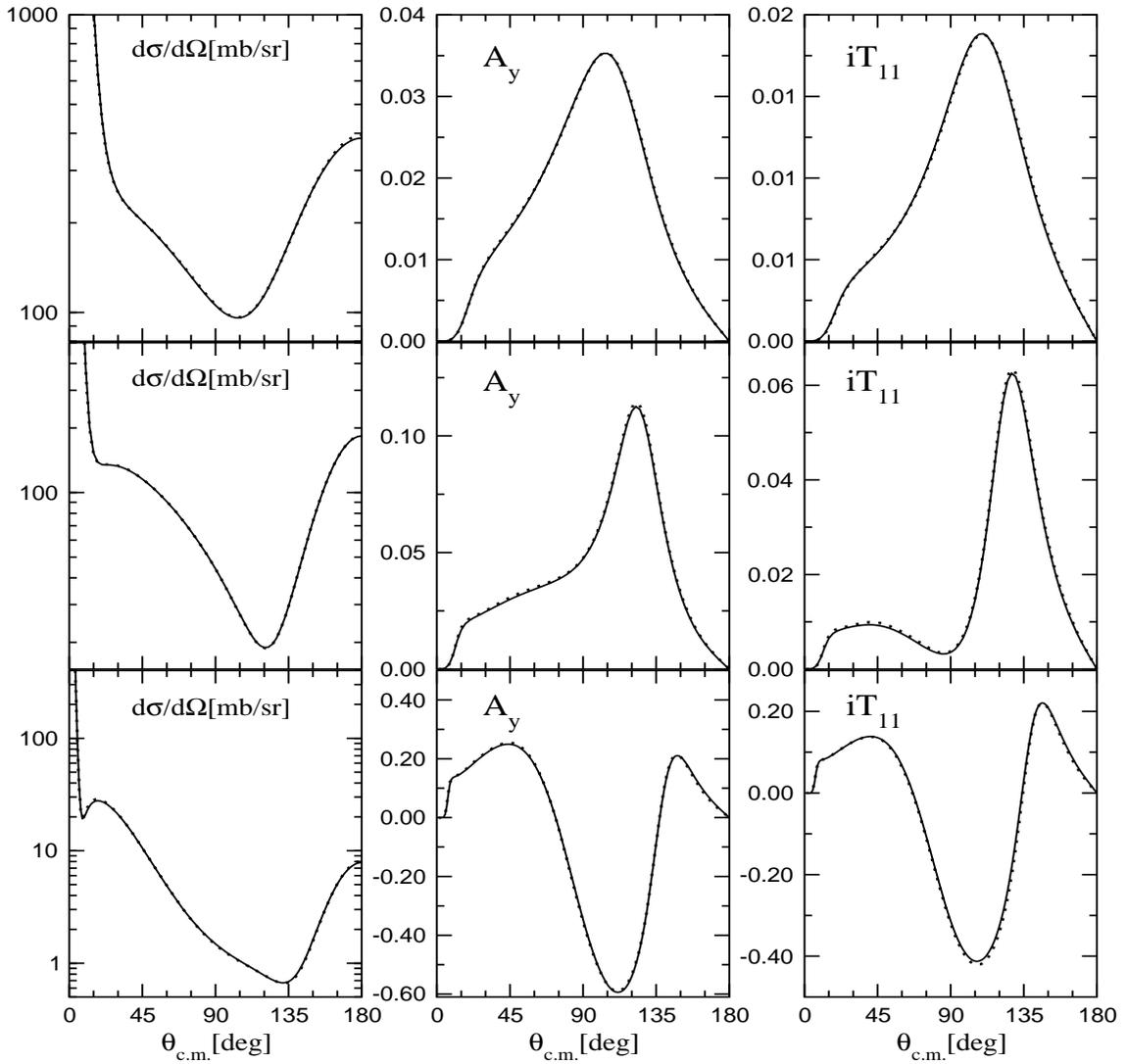}
\end{center}
\caption{\label{fig:obs1}
Differential cross section, and the proton and deuteron analyzing powers 
$A_y$ and $iT_{11}$ are shown for $pd$ elastic scattering at $3$ MeV 
(first row), $10$ MeV (second row) and $65$ MeV (third row) lab energies 
as functions of the c.m. scattering angle. 
Results obtained using KVP (thin solid line) and integral equation
approach (dotted line) are compared.}
\end{figure*}

\begin{figure*}
\begin{center}
\includegraphics[width=17cm,height=19cm]{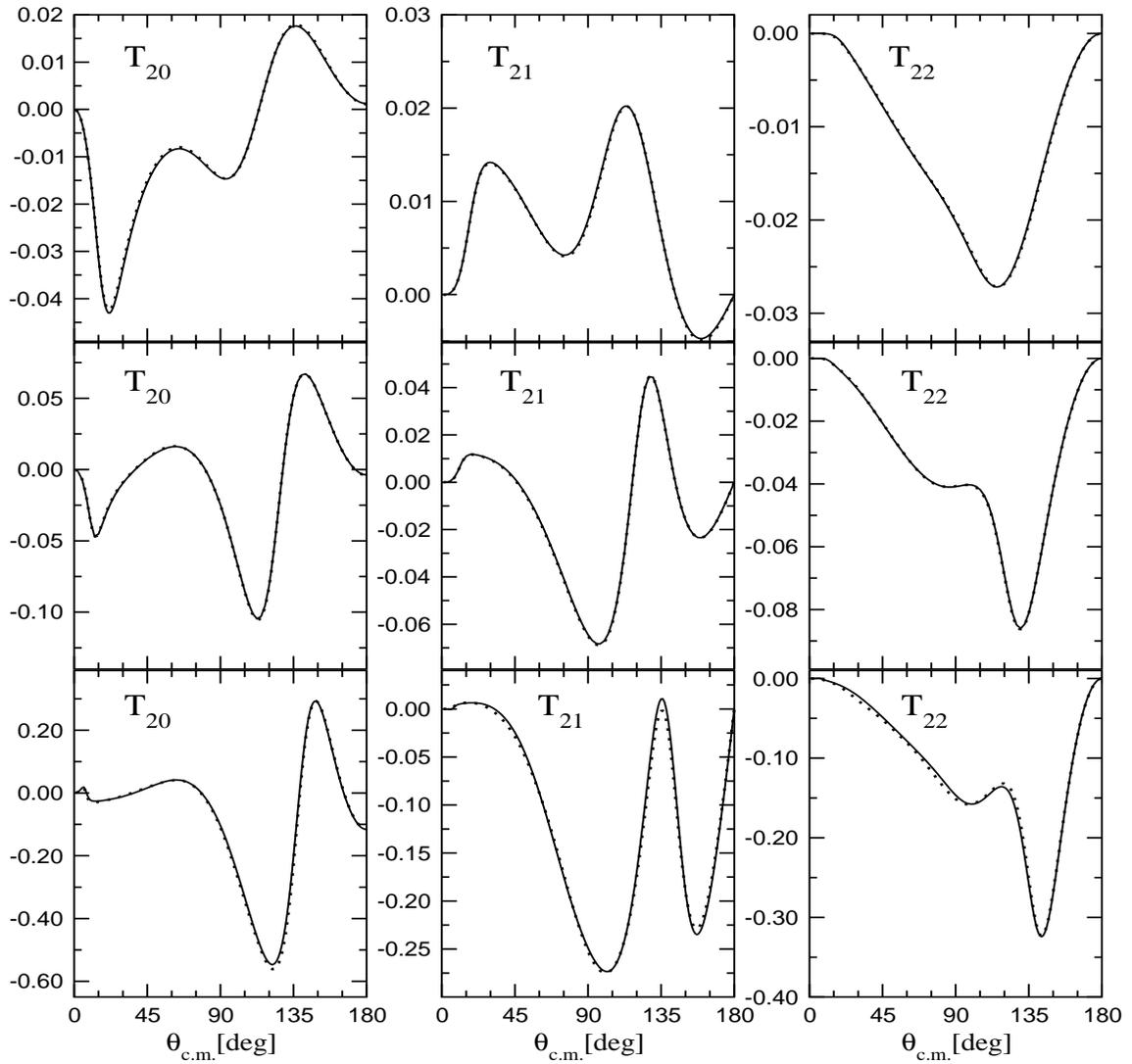}
\end{center}
\caption{\label{fig:obs2}
Tensor analyzing powers $T_{20}$, $T_{21}$, and $T_{22}$ are shown for 
$pd$ elastic scattering at $3$ MeV (first row), $10$ MeV (second row) 
and $65$ MeV (third row) lab energies as functions of the c.m. 
scattering angle. 
Results obtained using KVP (thin solid line) and integral equation
approach (dotted line) are compared.}
\end{figure*}

\begin{figure*}
\begin{center}
\includegraphics[width=17cm,height=19cm]{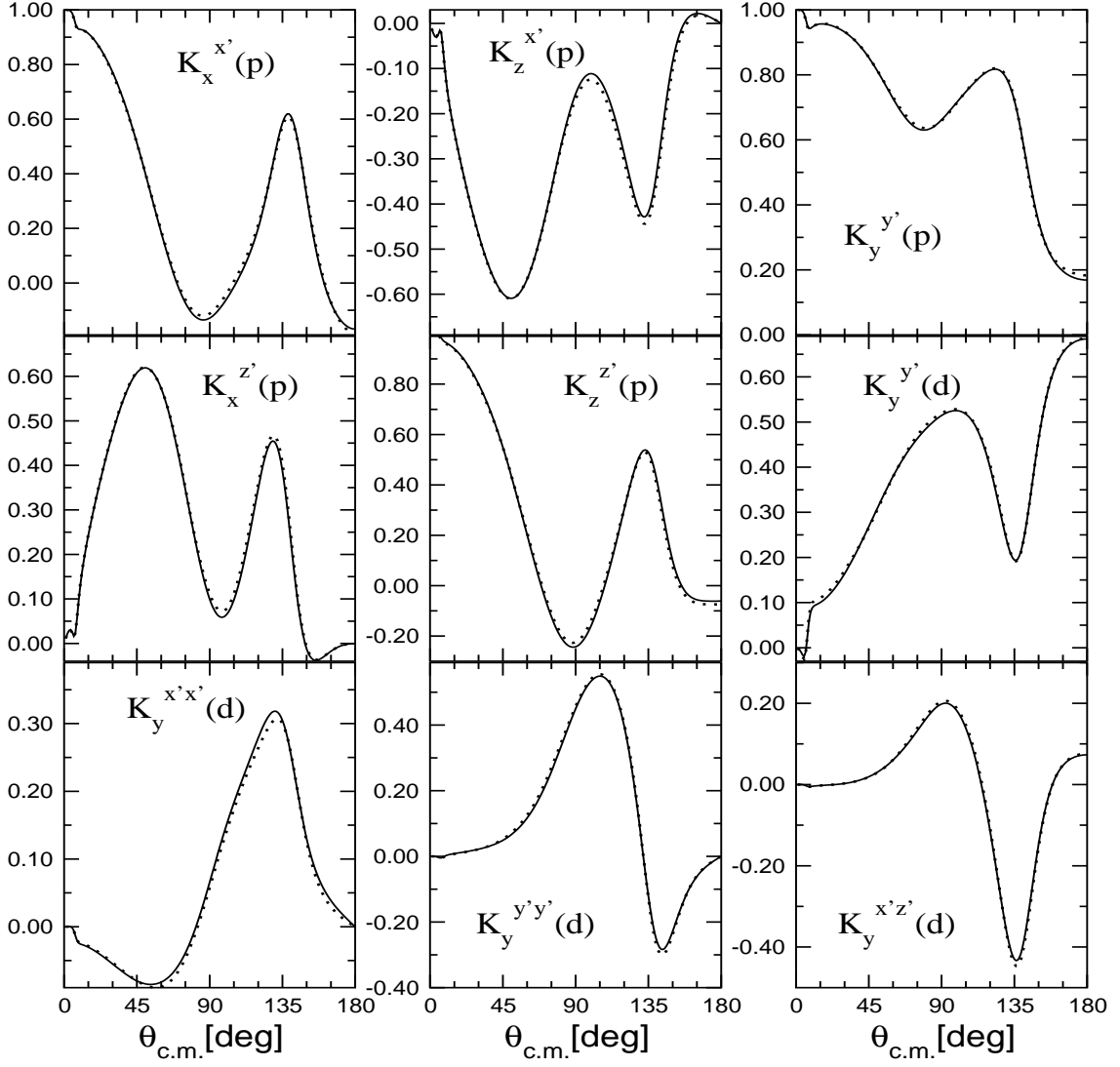}
\end{center}
\caption{\label{fig:stc}
A selection of proton to proton and proton to deuteron spin
transfer coefficients for $pd$ elastic scattering at $65$ MeV 
lab energy as function of the c.m.  scattering angle. 
Results obtained using KVP (thin solid line) and integral equation
approach (dotted line) are compared.}
\end{figure*}

In Figs.~\ref{fig:obs1} and  \ref{fig:obs2} 
we compare the differential cross section and vector and tensor
analyzing powers for $pd$ elastic scattering at the three selected 
energies, $3$, $10$ and $65$ MeV proton lab energies. 
In Fig.~\ref{fig:stc} a selection of spin transfer coefficients at
$65$ MeV is shown. In the figures, two different
curves are shown corresponding to calculations using the KVP
(thin solid line) and integral equation approach (dotted line).
By inspection of the figures one may conclude that the agreement is excellent 
because the numerical calculations agree to better than 1\%. 
In fact the curves are practically one on top of the other, the exceptions 
being the maximum of $T_{21}$ and some spin transfer coefficients
at $65$ MeV in which a small disagreement is observed.
Nevertheless, it is important to mention that in all cases the difference
between the two curves is smaller than the experimental accuracy
for the corresponding data sets. Likewise the agreement between the 
two calculations largely exceeds the agreement of any of them with
the data as shown in \Refs~\cite{kievsky:01a,deltuva:05a}.

\begin{figure*}
\begin{center}
\includegraphics[width=17cm,height=19cm]{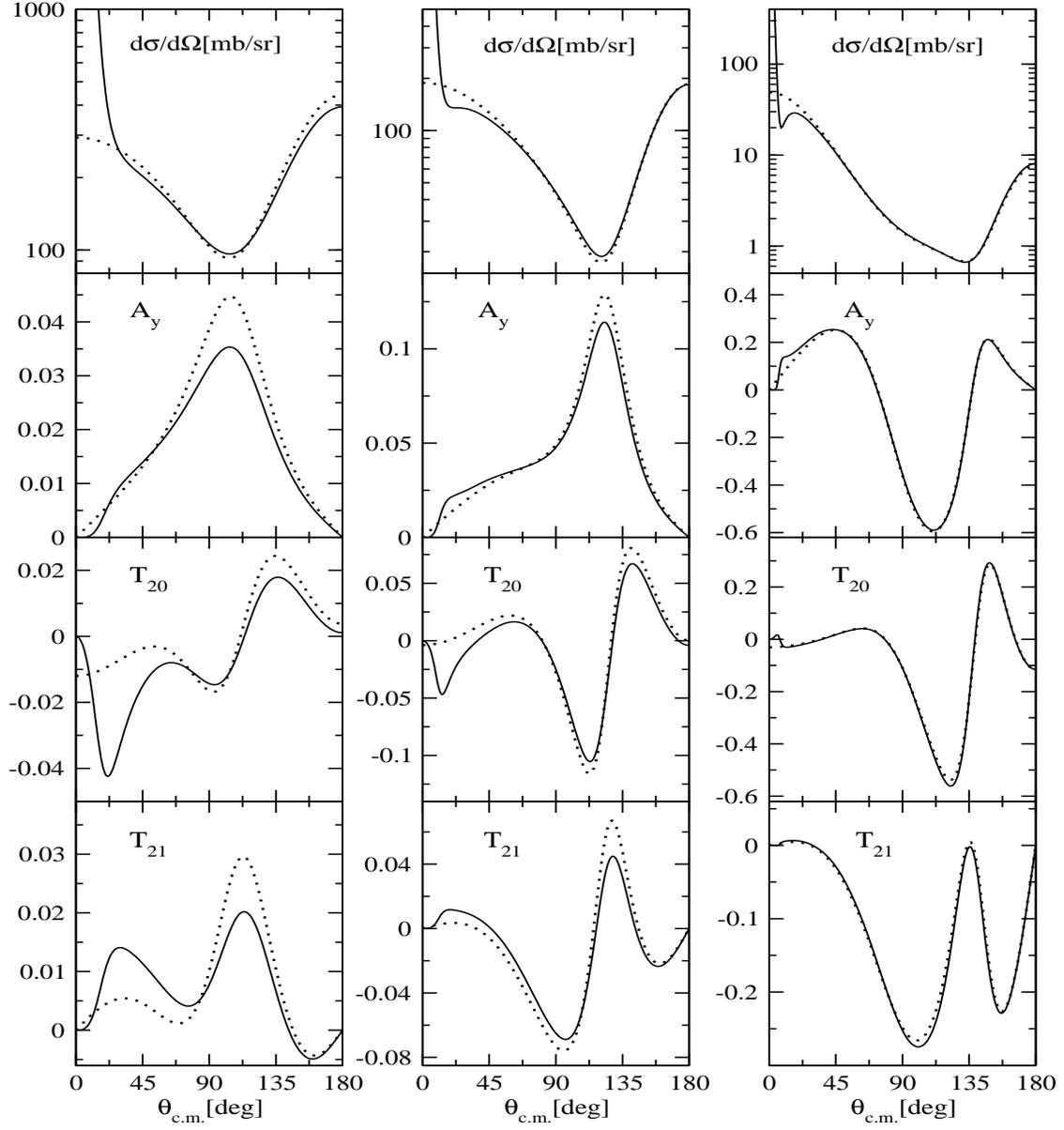}
\end{center}
\caption{\label{fig:obs3}
Calculations for $pd$ (solid line)  and $nd$ (dotted line) scattering
are compared for differential cross section, $A_y$, $T_{20}$ and
$T_{21}$ at $3$ MeV (first column), $10$ MeV (middle column) and
$65$ MeV (last column) nucleon lab energies.}
\end{figure*}

The present results can be used to study Coulomb effects by comparing
$nd$ to $pd$ calculations. 
In Fig.~\ref{fig:obs3} we analyze the evolution of the Coulomb effects
for the differential cross section, the nucleon analyzing 
power $A_y$ and two tensor analyzing powers, $T_{20}$ and $T_{21}$ at 
$3$, $10$ and $65$ MeV proton lab energies. 
In order to reduce the number of curves in the figure for the sake of clarity
we present results obtained using the integral equation approach. The results
obtained using the KVP for $nd$ scattering agree at the same level 
already shown for the $pd$ case in the previous figures. 
In Fig.~\ref{fig:obs3} the thin solid line denotes the $pd$ calculation 
whereas the dotted line denotes the corresponding $nd$ calculation. 
The latter agrees well with 
the results of other existing $nd$ calculations~\cite{witala:pc}.
From the figure we observe that Coulomb effects are appreciable at 
$3$ and $10$ MeV but are considerably reduced at $65$ MeV.
A more exhaustive analysis on Coulomb effects can be found
in Refs.~\cite{kievsky:01a,KRV01,deltuva:05a}.

In addition to the benchmark comparison using AV18 potential
we also give one result for the Malfliet-Tjon (MT) I-III potential,
in order to resolve an existing problem.
Reference \cite{suslov:04a} reports a disagreement between
$pd$ phase shifts results for MT I-III potential
calculated using the first technique of this paper,
the KVP \cite{kievsky:01a},
and the configuration-space Faddeev equations \cite{suslov:04a}.
The calculation based on the second technique of this paper,
the momentum-space integral equations \cite{deltuva:05a},
clearly confirms the results of \Ref~\cite{kievsky:01a}.
A detailed comparison of $pd$ and $nd$ phase shift results for
MT I-III potential is given in Table~\ref{tab:MT}.

In the following we discuss some of the limitations inherent to the
two methods used to describe $pd$ elastic scattering.
The KVP, as presented here, reduces the scattering problem to the solution
of a linear set of equations in which the matrix elements of
the Hamiltonian have to be computed between basis states; increasing the
energy, appreciable contributions from states with high values of 
$\ell_\alpha + L_\alpha$ appear. In order to take into account these
contributions, a very large basis has to be used with the consequence
that numerical instabilities start to appear.
In the integral equation approach at very low energies 
convergence in terms of screening radius requires
$R > 30 \fm$, which in turn increases the number of two-body partial waves
that are needed for convergence. The interplay of these two requirements
makes the integral equation solution unstable at those very low energies.
An interesting heuristic argument to understand the size of the screening
radius needed for convergence is the wave length $\lambda$ corresponding to
the on-shell momentum. At 3 MeV, 10 MeV, and 65 MeV proton lab energy,
for which a screening radius of 20 fm, 10 fm, and 7 fm is needed
for convergence,
$\lambda$ is  $24.8\fm$, $13.6\fm$, and $5.3\fm$, respectively.
It appears that for the calculation of $pd$ elastic scattering observables 
the screening has to be only so large that one wave
length can be accommodated in the Coulomb tail outside the range of the
hadronic interaction; seeing proper Coulomb over one wave length appears
enough  to provide, with the additional help of renormalization, the true
Coulomb characteristics of scattering despite screening.

\section{Conclusions \label{sec:conclusions}}
\bigskip 

In the present paper two methods devised to describe elastic
$pd$ scattering are compared for a wide range of energies. One of
the methods, the KVP, was developed a few years ago and
used to study how realistic potential models,
including two-body and three-body forces, describe the
elastic observables measured for that reaction. On the other
hand, numerical accurate results have been recently obtained solving
the AGS equation for $pd$ scattering using a screened Coulomb potential 
corrected for the unscreened limit using a renormalization
prescription. As has been briefly described in the present paper,
both methods are substantially different. It is satisfactory to observe 
that both methods produce essentially the same results for a large 
variety of elastic observables using a realistic two-nucleon potential.
We stress the fact that the selection of observables here presented is only
part of the observables compared. In all cases, similar patterns have
been obtained.  
In addition, by comparing the $pd$ calculations to the corresponding $nd$ 
calculations, Coulomb effects have been estimated. As expected these
effects are sizable at low energies but at the highest analyzed
energy, $65$ MeV, they are small, except at forward scattering angles.
From these considerations it is
possible to identify on a firm basis which $pd$ observables may or may
not be analyzed by calculations in which the Coulomb interaction
has been neglected.

We can conclude that at present it is possible to describe $pd$ elastic
scattering, including the Coulomb repulsion, using standard techniques
as the Faddeev equations in configuration and momentum space or
variational principles. Moreover, in Ref.~\cite{KVM04} the treatment
of other terms of the $NN$ electromagnetic potential as the magnetic
moment interaction has been discussed.

\begin{acknowledgments}
The authors are grateful to H.~Wita{\l}a for the comparison of $nd$ results.
A.D. is supported by the FCT grant SFRH/BPD/14801/2003,
A.C.F. in part by the grant POCTI/FNU/37280/2001, 
and P.U.S. in part by the DFG grant Sa 247/25.
\end{acknowledgments}

\bibliographystyle{prsty}


\newpage

\begin{table}[htbp]
  \centering
\begin{ruledtabular}
  \begin{tabular}{{l}*{4}{c}}
 & ${}^2 \delta_0$ & ${}^2 \eta_0$ & ${}^4 \delta_0$ & ${}^4 \eta_0$ \\ \hline
$nd$ at 14.1 MeV 
 & 105.48 & 0.4649 & 68.95 & 0.9782 \\
 & 105.49 & 0.4649 & 68.95 & 0.9782 \\
\hline
$nd$ at 42.0 MeV 
& 41.34 & 0.5022 & 37.72 & 0.9033 \\
& 41.36 & 0.5022 & 37.71 & 0.9033 \\
\hline
$pd$ at 14.1 MeV 
 & 108.44 & 0.4984 & 72.60 & 0.9795 \\
 & 108.39 & 0.4983 & 72.62 & 0.9796 \\
\hline
$pd$ at 42.0 MeV 
& 43.67 & 0.5056 & 39.95 & 0.9046 \\
& 43.70 & 0.5056 & 39.97 & 0.9046 \\
  \end{tabular}
\end{ruledtabular}
  \caption{ \label{tab:MT}
$nd$ and $pd$ phase-shift and inelasticity parameters
calculated with MT I-III potential. For each reaction,
the KVP and integral equation approach results are given 
in the first and second row, respectively.}
\end{table}

\newpage

\end{document}